\documentclass[letterpaper, 10 pt, conference]{ieeeconf}  % Comment this line out
\IEEEoverridecommandlockouts                              % This command is only
\usepackage{import}
\usepackage{color}
\usepackage{graphicx}
\usepackage{colortbl}

\usepackage[utf8]{inputenc}
\usepackage[T1]{fontenc}
\usepackage{amssymb}
\usepackage{amsmath}
\usepackage{url}
\usepackage{multirow}

\title{\LARGE \bf
CNN Filter Learning from Drawn Markers for the Detection of Suggestive Signs of COVID-19 in CT Images
}

\author{Azael M. Sousa$^{1}$, Fabiano Reis$^{2}$, Rachel Zerbini$^{2}$,  Jo\~{a}o L. D. Comba$^{3}$ and Alexandre X. Falc\~{a}o$^{1}$% <-this % stops a space
\thanks{$^{1}$Laboratory of Image Data Science (LIDS), Institute of Computing, University of Campinas}%
\thanks{$^{2}$Faculty of Medical Sciences, University of Campinas}%
\thanks{$^{3}$Institute of Informatics, Federal University of Rio Grande do Sul}
}

\begin{document}

\maketitle
\thispagestyle{empty}
\pagestyle{empty}

%%%%%%%%%%%%%%%%%%%%%%%%%%%%%%%%%%%%%%%%%%%%%%%%%%%%%%%%%%%%%%%%%%%%%%%%%%%%%%%%
\begin{abstract}

Early detection of COVID-19 is vital to control its spread. Deep learning methods have been presented to detect suggestive signs of COVID-19 from chest CT images. However, due to the novelty of the disease, annotated volumetric data are scarce. Here we propose a method that does not require either large annotated datasets or backpropagation to estimate the filters of a convolutional neural network (CNN). For a few CT images, the user draws markers at representative normal and abnormal regions. The method generates a feature extractor composed of a sequence of convolutional layers, whose kernels are specialized in enhancing regions similar to the marked ones, and the decision layer of our CNN is a support vector machine. As we have no control over the CT image acquisition, we also propose an intensity standardization approach. Our method can achieve  mean accuracy and kappa values of $0.97$ and $0.93$, respectively, on a dataset with 117 CT images extracted from different sites, surpassing its counterpart in all scenarios.

\end{abstract}
%%%%%%%%%%%%%%%%%%%%%%%%%%%%%%%%%%%%%%%%%%%%%%%%%%%%%%%%%%%%%%%%%%%%%%%%%%%%%%%%

\section{INTRODUCTION}
\label{sec:introduction}

The world is currently facing a public health crisis caused by COVID-19, an infectious disease named after the new coronavirus SARS-CoV-2. According to the World Health Organization (WHO), over $100$ million people have been infected worldwide by COVID-19. An early diagnosis can significantly reduce the transmission rate by keeping the infected under lockdown. Although the reverse-transcriptase polymerase chain reaction (RT-PCR) is considered the gold-standard for the diagnosis of COVID-19, computed tomography (CT) images are also important to support the diagnosis, monitor the progression of the disease, and detect possible complications~\cite{fang2020sensitivity,ng2020imaging,rosa2020covid}. 
Studies have reported cases with negative RT-PCR and CT exams with abnormal characteristics that became positive RT-PCR later on~\cite{xie2020chest,huang2020use}. As consequence, laboratories are employing both RT-PCR and chest CT for patients with suspect of COVID-19 infection~\cite{ng2020imaging}.

%Chest CT produces high-resolution images with moderate contrast among organs, muscles, tissues, and bones. A CT is composed of a stack of sequential slices, forming a volumetric image. From this CT image it is possible to extract relevant information that can be used to determine whether there is an abnormal manifestation as well as the location and spatial extension of the illness~\cite{silva2020covid}.

\begin{figure}[ht!]
    \centering
    \includegraphics[width=0.45\textwidth]{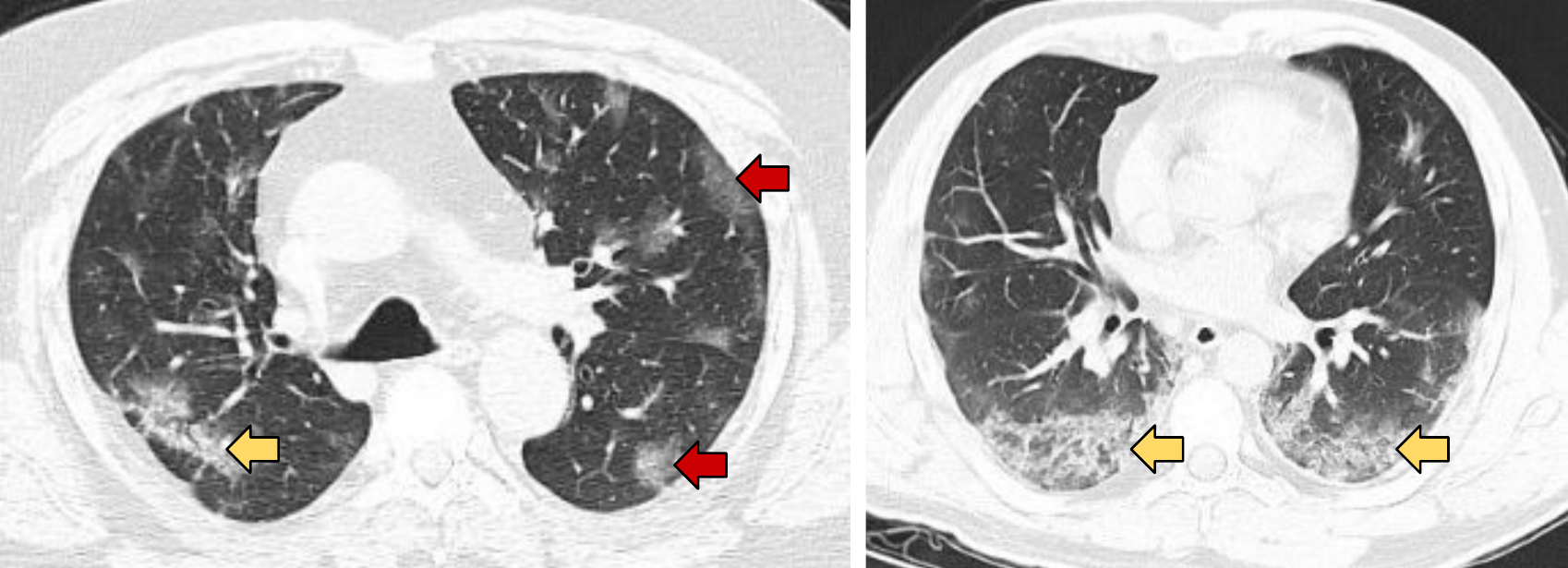}
    \caption{CT images from patients with confirmed diagnosis of COVID-19. The red arrows indicate ground-glass opacities whereas yellow arrows indicate consolidations.}
    \label{fig:exmaple_covid}
\end{figure}

In CT images, common findings of COVID-19 include (i) peripheral, bilateral, ground-glass opacities (GGO) with or without consolidation, (ii) multifocal GGO of rounded morphology, and (iii) reverse halo sign or other findings of organizing pneumonia~\cite{kwee2020chest}, as shown in Figure~\ref{fig:exmaple_covid}. As these findings may appear in other pulmonary conditions, they can only be suggestive of COVID-19.

Due to the urgency of the problem, several deep learning methods, mostly Convolutional Neural Networks (CNN), have been presented to detect suggestive patterns of COVID-19~\cite{shi2020review,shi2020large,silva2020covid,polsinelli2020light,jin2020development,chen2020deep,jin2020ai}. Wang \textit{et al.}~\cite{wang2020weakly} employ a 3D CNN, called DeCovNet, that can achieve an accuracy of approximately 90\% on an in-house dataset with $630$ CT images. Polsinelli \textit{et al.}~\cite{polsinelli2020light} uses a light design of the 2D SqueezeNet CNN. The authors report an accuracy of 83\% on two public datasets~\cite{zhao2020covid,sirm}. Silva \textit{et al.}~\cite{silva2020covid} propose a 2D CNN called EfficientCovidNet. The authors employ a voting mechanism that determines the patient's outcome based on the number of slices classified as COVID-19. This method can achieve an accuracy of 89\% when tested on new images of the same dataset used for training, but presents an accuracy drop to 56\% when tested on images from a different dataset. Other studies~\cite{jin2020development,chen2020deep,jin2020ai} have addressed the problem by first identifying lung lesions and then giving the diagnosis of the patient based on the classification of those lesions.
%Most of these methods fall into one of the following categories regarding the types of images to be classified: (i) COVID-19 and healthy images, (ii) COVID-19, other diseases and healthy images, and (iii) COVID-19 and Non-COVID-19. 

\begin{figure*}[ht!]
    \centering
    \includegraphics[width=0.99\textwidth]{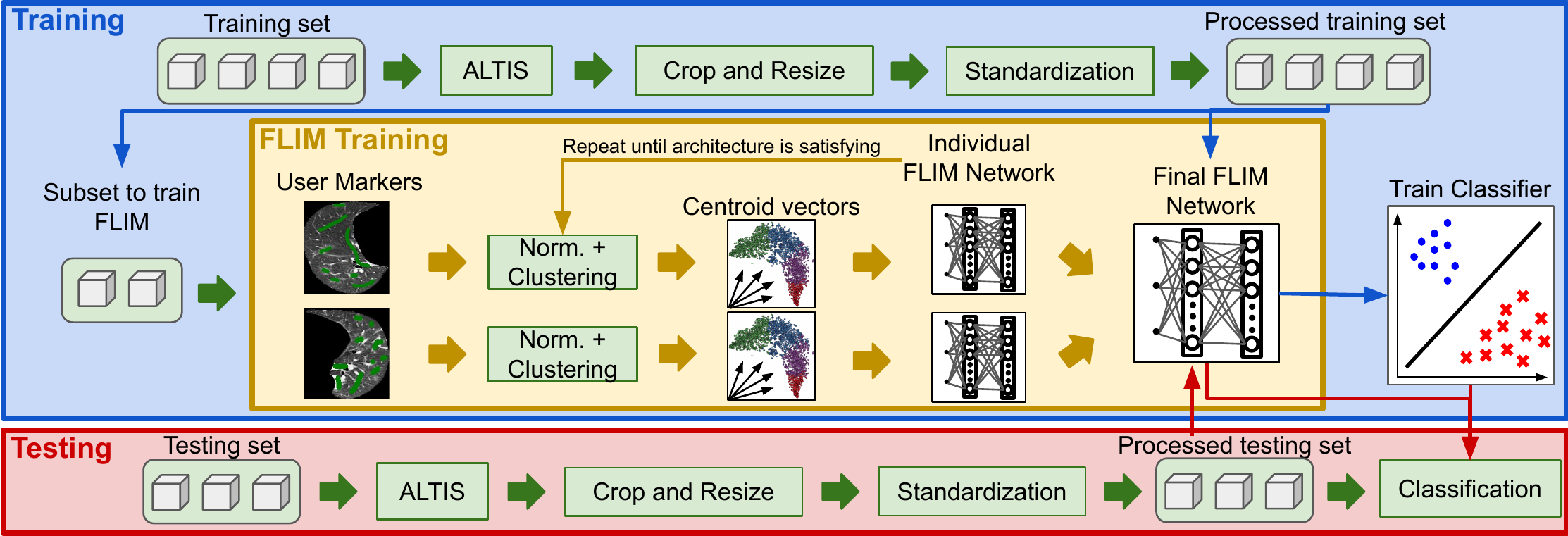}
    \caption{Pipeline of the proposed method. A subset with few images is selected from a preprocessed training set to train the FLIM network. After the FLIM training, the entire training set is passed through the network for feature extraction, and a SVM classifier is trained. The classification of the test set is performed over the activations of the FLIM network.}
    \label{fig:pipeline}
\end{figure*}

One of the main drawbacks of deep learning is the number of annotated images required for training. Typically, it requires large annotated datasets to learn representative features from different abstraction levels by backpropagation, making the trained model robust to real-world scenarios. However, such datasets can be challenging to gather. Due to its novelty, publicly available datasets of COVID-19 with standardized, annotated, and volumetric CT exams are scarce. Another obstacle faced when dealing with deep learning models is the architecture choice. Many methods employ off-the-shelf architectures for applications of entirely different domains, leading to poor computational resources management, too many training samples, and still non-explainable solutions. 

We involve the specialist in the training process to address the above issues. By taking advantage of the expert's knowledge about the problem, we can build a more efficient model with less trainable parameters and considerably reduced number of annotated training images. Our approach explores a recently proposed method for feature extraction (i.e., a network formed by a sequence of convolutional layers only), named \textit{Feature Learning from Image Markers} (FLIM)~\cite{de2020feature,de2020learning}. FLIM kernels are estimated based on a set of markers drawn by the user in very few training images (e.g., two or three images per class), with no backpropagation. These markers indicate pertinent image regions for feature extraction, creating kernels that enhance regions similar to the marked ones. 

In this work, we extend the 2D FLIM network to detect suggestive signs of COVID-19 in volumetric CT images. In addition to that, we adopt an interactive architecture learning approach to construct the FLIM architecture. The network designer incorporates his/her knowledge in the training loop by studying the impact that the selected hyperparameters have on the network's accuracy over a small validation set and adapting them accordingly.
%That is, the network designer incorporates his knowledge in the training loop by selecting the network hyperparameters that maximizes its performance.

In Section II we describe the proposed method, Section III depicts the experimental procedure and assessment of our method in comparison with another approach from the state-of-the-art, and Section IV states the conclusion of our work.
\section{METHOD}
\label{sec:method}

The proposed method is divided into four stages: lung segmentation, pre-processing, feature extraction, and classification. Figure~\ref{fig:pipeline} shows the entire pipeline.

\subsection{Lung Segmentation}
Image segmentation consists of defining the spatial extension of one or multiple objects of interest. The segmentation stage is of paramount importance in deep learning as it reduces the regions to be processed and consequently the computational effort. In this work, we employed the method \textit{ALTIS} (\textbf{A}utomatic \textbf{L}ungs and \textbf{T}rachea \textbf{I}mage \textbf{S}egmentation)~\cite{sousa2019altis} for the segmentation of the lungs. ALTIS is a graph-based method that explores the Image Foresting Transform (IFT) framework~\cite{falcao2004image} to estimate object seeds and delineate the lungs and trachea-bronchi as separated objects.

\subsection{Preprocessing}
Each lung from the CT image is first cropped according to the mask generated by ALTIS. To guarantee feature vectors of the same size, the cropped-lung images are resized to a lower common dimension of $200\times200\times200$ voxels; wherein ground-glass opacities can still be clearly seen.

\begin{figure}[ht!]
    \centering
    \includegraphics[width=0.45\textwidth]{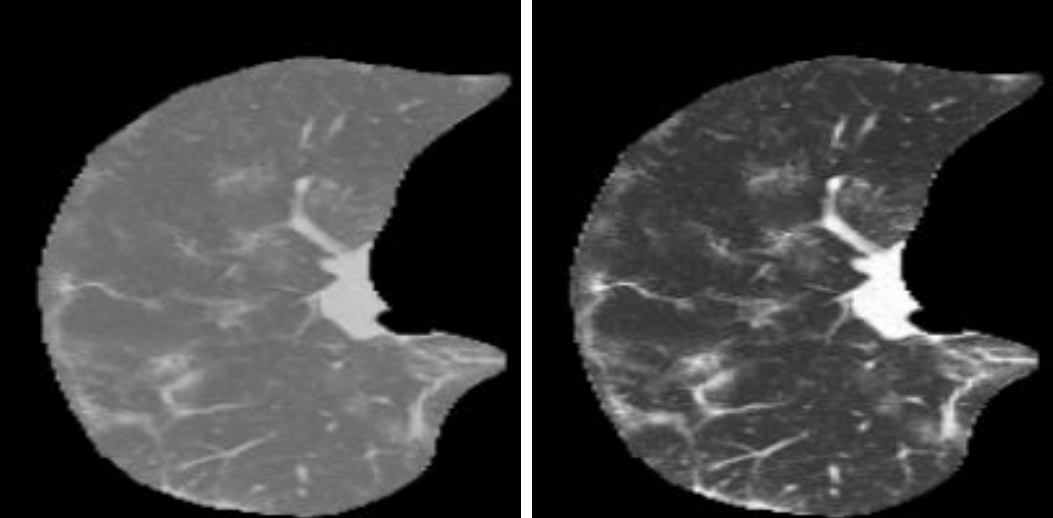}
    \caption{Standardization method. The left image is the original right lung cropped and resized, whereas the right one is the cropped and resized right lung image after standardization.}
    \label{fig:standardization}
\end{figure}

The lack of control during image acquisition and reconstruction creates a disparity problem regarding the intensity feature that may hampers deep learning methods~\cite{berenguer2018radiomics}. For that reason, we applied a standardization method to keep each lung image within a fixed intensity range. This method identifies the first two summits closest to each edge of the lung image's intensity histogram. The first summit represents the darker and typical parenchyma region, while the second one represents the bright veins/arteries and part of the mediastinal region. Then, the histogram is shifted to a previously established spectrum. Figure~\ref{fig:standardization} shows an example of the standardization method.
%according to the patient's needs}

%The versatility of CT exams allows the customization of image acquisition and reconstruction protocols according to the patient's needs~\cite{liang2019ganai,selim2020stan}. While it expands the scope of scenarios for CT usage, it creates a data discrepancy problem regarding radiomic features (\textit{e.g.} intensity, texture, and shape) that hampers deep learning methods~\cite{berenguer2018radiomics}. For that reason, we applied a standardization method to keep each image object within a fixed intensity range. This method identifies the first two summits closest to each edge of the lung image's intensity histogram. The first summit represents the darker and typical parenchyma region, while the second one represents the bright veins/arteries and part of the mediastinal region. Then, the histogram is shifted to a previously established spectrum. Figure~\ref{fig:standardization} shows an example of the standardization method.
%according to the patient's needs

\subsection{Feature Extraction}
Feature extraction is an essential part of any machine learning algorithm. CNN's have powerful feature extractors that can learn representative patterns from training samples. In our case, they consist of a sequence of convolutional layers, each containing marker-based normalization, convolution with a filter bank, ReLU activation and, optionally, a max-pooling operation. Let $\mathcal{F}_l$ be the set of convolutional filters (kernels) for a layer $l \in \{1,2,\dots,L\}$. The FLIM network estimates $\mathcal{F}_l$ by analyzing the markers drawn by the user, with no backpropagation. It works as an attention mechanism for relevant regions.

Let $\mathcal{D}$ be the training set containing CT images from both classes, with $m$ channels. The convolution operation of an image $I \in \mathcal{D}$ with a kernel $f \in \mathcal{F}_l$ highlights local patterns that matches $f$. It is denoted as $P_{I(v)} \ast f$, where $P_{I(v)}$ is a patch of size $h \times w \times d \times m$, extracted from the image $I$ around the voxel $v$. A unit-norm kernel $f$ may be seen as an orthogonal vector to a hyperplane at the origin of the patch feature space. The convolution with $f$ measures the distance to that hyperplane and the ReLU function eliminates regions with negative distances (non-activated regions). A kernel $f$ is said discriminative for a given class $c \in \{1,2,..., C\}$ when the regions activated by $f$ are different from those it activates for other classes. 

%the local patterns of a certain class $c \in \{1,2,...,C\}$ are on the positive side of its hyperplane, while the local patterns of the remaining ones are on the negative side.

Let $\mathcal{M_I}$ be the set of markers drawn by the user on image $I \in \mathcal{D}$ (green engravings on ``User Markers'' at Figure~\ref{fig:pipeline}). Every voxel $v \in \mathcal{M_I}$ has a patch $P_{I(v)}$ centered on it. Let $\mathcal{P}_I$ be the set of patches from image $I$. Marker-based normalization is needed to eliminate bias in the FLIM network. It shifts all patches by moving their geometric center to the origin of the patch feature space. By grouping the normalized patches $\mathcal{P}_I$ (e.g., using K-Means) such that $\mathcal{G}_I$ is the resulting set of groups, each unit vector pointing towards the centroid of each group $g \in \mathcal{G}_I$ is a potential kernel $f_g$ for $\mathcal{F}_l$ in the FLIM architecture. Therefore, the complete set of filters for layer $l$ is $\mathcal{F}_l = \cup_{g \in \mathcal{G}_I, I \in \mathcal{D}} f_g$.

%Let $\mathcal{P}_I$ be the set of patches from image $I$ and $\mathcal{G}_I$ the set of groups resulted from a clustering method (in this work, the K-Means) applied over $\mathcal{P}_I$. The unit vector pointing towards the centroid of each group $g \in \mathcal{G}_I$ is a potential kernel $f_g$ for $\mathcal{F}_l$ of the FLIM architecture. A marker-based normalization is applied over the patches to avoid the need to estimate the bias, \textit{i. e.} all patches are shifted so that the mean of $\mathcal{P}_I$ coincides with the origin. Therefore, the complete set of filters for layer $l$ is $\mathcal{F}_l = \cup_{g \in \mathcal{G}_I, I \in \mathcal{D}} f_g$.

%\begin{equation}
%    \mathcal{F}_l = \bigcup_{\substack{g \in \mathcal{G}_I \\ I \in \mathcal{D}}} f_g
%\end{equation}

To remove redundant kernels from $\mathcal{F}_l$, whenever necessary, we use a methodology similar to the one described in~\cite{sousa2020approach}. The idea is to make all kernels linearly independent. For that reason, the Principal Component Analysis (PCA) method is applied over $\mathcal{F}_l$ generating a set of orthogonal eigenvectors and eigenvalues. The $k_l$ eigenvectors associated with the highest eigenvalues are selected as the final set of kernels, where $k_l$ is the number of desired kernels for layer $l$. The kernel estimation procedure is repeated for every subsequent layer $l+1$ by extracting patches from marker voxels at the output of layer $l$, until the network designer is satisfied with the architecture. However, the user only draws markers over the original images (at the input layer). 

\subsubsection{Interactive Architecture Learning}
To define the FLIM architecture, we explore the maximum class separability that a layer can achieve on its own. For that purpose, each layer is built individually by the network designer. Its hyperparameters, \textit{e. g.} number of kernels, kernel size, and dilation rate (in case of atrous convolution) are intuitively optimized according to the model's effectiveness on a small validation set. Once a layer has reached its maximum separability, the network designer decides to either build another layer and repeat the process or finish the model.

\subsection{Classification}
Once the network architecture and respective weights have been computed, all training data, both normal CT and with COVID-19, are passed forward through the network. The output of the last convolutional layer is then used to train a Support Vector Machine (SVM) classifier~\cite{Cortes1995support}.
%, one-versus-one
\section{EXPERIMENTS AND DISCUSSION}
\label{sec:experiments}

% \usepackage{multirow}
\begin{table*}[ht!]
\caption{Accuracy (Acc.) and Kappa scores for the classification of COVID-19}
\label{tab:results}
\centering
\begin{tabular}{|l|l|l|l|l|l|l|l|l|l|l|l|l|}
\hline
\multicolumn{1}{|c|}{}                         & \multicolumn{2}{c|}{Split 1} & \multicolumn{2}{c|}{Split 2} & \multicolumn{2}{c|}{Split 3} & \multicolumn{2}{c|}{Split 4} & \multicolumn{2}{c|}{Split 5} & \multicolumn{2}{c|}{Mean and Stdev}                  \\ \cline{2-13} 
\multicolumn{1}{|c|}{\multirow{-2}{*}{Method}} & Acc.       & Kappa       & Acc.       & Kappa       & Acc.       & Kappa       & Acc.       & Kappa       & Acc.       & Kappa       & Acc.                       & Kappa     \\ \hline
deCovNet$\ast$                                       & 0.87           & 0.73        & 0.84           & 0.69        & 0.87           & 0.74        & 0.92           & 0.83  & 0.94        & 0.89        & $0.89 \pm 0.04$                      & $0.77 \pm 0.08$ \\ \hline
FLIM layer 1                                   & 0.95           & 0.90        & 0.94           & 0.87        & \textbf{0.96}  & \textbf{0.92}        & \textbf{0.95} & \textbf{0.90}        & 0.91           & 0.81        & $0.94 \pm 0.02$ & $0.88 \pm 0.04$ \\ \hline
FLIM layer 2                                   & \textbf{0.96}  & \textbf{0.92} & \textbf{0.97} & \textbf{0.95} & \textbf{0.96} & \textbf{0.92}      & \textbf{0.95} & 0.89        & \textbf{0.97}           & \textbf{0.95} & $\mathbf{0.97 \pm 0.01}$                      & $\mathbf{0.93 \pm 0.02}$ \\ \hline
\end{tabular}
\end{table*}

\iffalse
% \usepackage{multirow}
\begin{table*}[ht!]
\caption{Accuracy (Acc.) and Kappa scores for the classification of COVID-19}
\label{tab:results}
\centering
\begin{tabular}{|l|l|l|l|l|l|l|l|l|l|l|l|l|}
\hline
\multicolumn{1}{|c|}{\multirow{2}{*}{Method}} & \multicolumn{2}{c|}{Split 1} & \multicolumn{2}{c|}{Split 2} & \multicolumn{2}{c|}{Split 3} & \multicolumn{2}{c|}{Split 4} & \multicolumn{2}{c|}{Split 5} & \multicolumn{2}{c|}{Mean}    \\ \cline{2-13} 
\multicolumn{1}{|c|}{}                        & Acc.       & Kappa       & Acc.       & Kappa       & Acc.       & Kappa       & Acc.       & Kappa       & Acc.       & Kappa       & Acc.      & Kappa       \\ \hline
deCovNet                                      & 0.87        & 0.73       & 0.84        & 0.69      & 0.87        & 0.74       & 0.92        & \textbf{0.83}       & \textbf{0.94}        & \textbf{0.89}       & $0.89 \pm 0.04$  & $0.77 \pm 0.08$ \\ \hline
Ours                                          & \textbf{0.93}       & \textbf{0.87}        & \textbf{0.93}       & \textbf{0.84}        & \textbf{0.95}       & \textbf{0.90}       & 0.92       & 0.82       & 0.87           & 0.70       & $\mathbf{0.92 \pm 0.03}$ & $\mathbf{0.82 \pm 0.07}$ \\ \hline
\end{tabular}
\end{table*}
\fi

\subsection{Datasets}
\label{ssec:datasets}
The Faculty of Medical Sciences at UNICAMP, Brazil, has provided an annotated dataset containing 66 volumetric CT images of patients with RT-PCR positive for COVID-19. Additionally, we have gathered 51 volumetric CT images with normal parenchyma from different sites. All images were interpolated to an isotropic voxel dimension of $1\times1\times1$ mm$^3$. We conducted experiments on five random stratified splits where 50\% were used to train the SVM classifier and 50\% for testing. From the SVM training set of every split, 10\% of the images with COVID-19 (\textit{i.e.} about 6 images) were separated to train the FLIM network (i.e., 3 images for marker selection and 3 images for hyperparameter selection based on validation, see Section~\ref{ssec:flim_architecture}). The splits were performed in a patient-wise manner to ensure learning integrity, where images from the same patient are either in the training set or test set but never in both.

\subsection{FLIM architecture}
\label{ssec:flim_architecture}
The FLIM architecture was determined by analyzing its accuracy and kappa scores on the validation split. To avoid skewing the interactive architecture learning process, we selected four different sets of images to train FLIM, extracted from the training/validation split. The final model was the one with the highest mean accuracy and kappa over all four sets. It has two convolutional layers with two kernels of size $3 \times 3 \times 3$, dilation rate of $3$, and pooling with a stride of $4$. The ReLU activation function was used in all layers. 
%No batch normalization was employed as we noted to be harmful to the FLIM network learning.

\subsection{Compared Method}
\label{ssec:compared_method}
The method used for comparison is known as deCoVNet~\cite{wang2020weakly}. It is a 3D deep CNN composed of a convolutional layer with a kernel of size $5 \times 7 \times 7$, batch normalization and pooling, followed by 2 3D residual blocks. The last part is a progressive classifier with 3 convolutional layers and a fully connected one with the softmax function. Henceforth we call deCoVNet$\ast$ the deCoVNet model trained from scratch with the same 50\%-50\% data splits described in Section~\ref{ssec:datasets}.

\subsection{Results and Discussion}

The accuracy and kappa scores of the proposed method and the deCoVNet$\ast$ for the classification of COVID-19 patterns are presented in Table~\ref{tab:results}. Note that our method surpassed the deCovNet$\ast$ in all five splits by a considerable margin. The data imbalance between the classes makes kappa more sensitive than accuracy. The inferior results achieved by deCovNet$\ast$ might be due to two reasons: (i) the size of the dataset available in this study is five times smaller than the in-house dataset used in the deCovNet's original paper~\cite{wang2020weakly}, and (ii) the number of training samples (about 58 images) is considerably lower than the number used in the original paper. As a result, backpropagation might have been affected by overfitting. Since our method is not restrained to backpropagation, we successfully trained the FLIM network with a small set of CT images, generating more representative features to train the SVM classifier.
%the amount of data separated for training. In the deCovNet's original paper, the authors use a division of $80\%$ for training and $20\%$ for testing, while it this study, we reduce the training set to $50\%$ and increased the testing set $50\%$ as well.
%the proportion of data separated for testing and training in this study. Compared to the one used train the deCovNet in the original paper ($80\%$ for training and $10\%$ for testing), the backpropagation may have gotten jeopardized by the lack of sufficient samples. Since our method is not restrained by backpropagation, we successfully trained the FLIM network with a small set of CT images, generating representative features for the SVM classifier.

The standardization method plays an essential role in the entire process. Without it, marker-based normalization does not work properly and the performances of the FLIM network and SVM are seriously affected.  That is, images with normal parenchyma might present  activated regions by kernels specialized in enhancing ground-glass opacities. 
\section{Conclusion}

In this work, we extended from 2D to 3D a feature extraction method, named FLIM, based on image markers and a sequence of convolutional layers. Together with a SVM classifier, the method was used to detect suggestive signs of COVID-19 in CT images. The proposed method is backpropagation-free and computes its kernel weights by analyzing markers drawn by the user on very few images. The whole process requires segmenting the lungs and trachea from CT exams, preprocess the images, train a FLIM network for feature extraction, and train a SVM classifier for the detection of abnormal images with ground-glass opacities. The experiments showed that the proposed method can surpass deCovNet*, a state-of-the-art approach, suggesting that FLIM could generate more representative image features for the given problem. We intend to make our model available for public use and design another FLIM network for ground-glass segmentation.
%and evaluation of the compromised volume of the lungs.
%pursue the work towards the
\section{Acknowledgements}

The authors thank the financial support from CAPES, CNPq (303808/2018-7 grant) and FAPESP (2014/12236-1 and 2017/03940-5 grants).

\bibliographystyle{IEEEtranS}
\bibliography{refs}

\end{document}